\documentclass[9pt,twocolumn,twoside]{gsajnl}

\newcommand{\absdiv}[1]{
  \par\addvspace{.5\baselineskip}
  \noindent\textbf{#1}\quad\ignorespaces
}

\usepackage{epstopdf}
\usepackage{multirow}
\usepackage{float}
\usepackage[super]{natbib}
\usepackage{hyperref}
\hypersetup{
    colorlinks=true,
    linkcolor=blue,
    filecolor=magenta,      
    urlcolor=cyan,
    pdftitle={Overleaf Example},
    pdfpagemode=FullScreen,
    }
\usepackage{caption}
\usepackage{subcaption}
\usepackage{stfloats}
\usepackage[figuresright]{rotating}
\usepackage{pdflscape}
{\clearpage\pagebreak[4]\global\pdfpageattr\expandafter{\the\pdfpageattr/Rotate 90}}%
{\clearpage\pagebreak[4]\global\pdfpageattr\expandafter{\the\pdfpageattr/Rotate 0}}%

\renewenvironment{thebibliography}[1]{
  \begin{oldthebibliography}{#1}
    \setlength{\itemsep}{0em}
    \setlength{\parskip}{0em}
}
{
  \end{oldthebibliography}
}

\runningtitle{The choroid in a kidney donor and recipient cohort}
\runningauthor{Burke \textit{et al.}}

\title{Evaluation of an automated choroid segmentation algorithm in a
longitudinal kidney donor and recipient cohort}

\author[1,$\ast$]{Jamie Burke}
\author[2]{Dan Pugh}
\author[2]{Tariq Farrah}
\author[3]{Charlene Hamid}
\author[4]{Emily Godden}
\author[5]{Tom MacGillivray}
\author[2]{Neeraj Dhaun}
\author[6]{J. Kenneth Baillie}
\author[1]{Stuart King}
\author[7, 8]{Ian J.C. MacCormick}

\affil[1]{School of Mathematics, College of Science and Engineering, University of Edinburgh, Edinburgh, UK}
\affil[2]{British Heart Foundation Centre for Cardiovascular Science, University of Edinburgh, Edinburgh, UK}
\affil[3]{Imaging Facility, University of Edinburgh, The Queen’s Medical Research Institute, Edinburgh, UK}
\affil[4]{Emergency Department, Royal Infirmary of Edinburgh, Edinburgh, UK}
\affil[5]{Centre for Clinical Brain Sciences, University of Edinburgh, Edinburgh, UK}
\affil[6]{Deanery of Clinical Sciences, College of Medicine and Veterinary Medicine, University of Edinburgh, Edinburgh, UK}
\affil[7]{Centre for Inflammation Research, The Queen’s Medical Research Institute, University of Edinburgh, Edinburgh, UK}
\affil[8]{Institute for Adaptive and Neural Computation, School of Informatics, University of Edinburgh, Edinburgh, UK}

\correspondingauthoraffiliation[$\ast$]{Corresponding and lead author; \\ Email address: Jamie.Burke@ed.ac.uk; \\ Word count: 4589; \\ Funding information: Medical Research Council [grant number MR/N013166/1]; \\ Commercial relationships: \textbf{Jamie Burke}, None; \textbf{Dan Pugh}, None; \textbf{Tariq Farrah}, None; \textbf{Charlene Hamid}, None; \textbf{Emily Godden}, None; \textbf{Tom MacGillivray}, None; \textbf{Neeraj Dhaun}, None; \textbf{J. Kenneth Baillie}, None; \textbf{Stuart King}, None; \textbf{Ian J.C. MacCormick}, None.}

\begin{abstract}

\absdiv{Purpose} 
To evaluate the performance of an automated choroid segmentation algorithm in optical coherence tomography (OCT) data using a longitudinal kidney donor and recipient cohort.
\absdiv{Methods}
We assessed 22 donors and 23 patients requiring renal transplantation over up to 1 year post-transplant. We measured choroidal thickness (CT) and area and compared our automated CT measurements to manual ones at the same locations. We estimated associations between choroidal measurements and markers of renal function (estimated glomerular filtration rate (eGFR), serum creatinine and urea) using correlation and linear mixed-effects (LME) modelling.
\absdiv{Results} 
There was good agreement between manual and automated CT. Automated measures were more precise because of smaller measurement error over time. External adjudication of major discrepancies were in favour of automated measures. Significant differences were observed in the choroid pre- and post-transplant in both cohorts, and LME modelling revealed significant linear associations observed between choroidal measures and renal function in recipients. Significant associations were mostly stronger with automated CT (eGFR \textit{P}<0.001, creatinine \textit{P}=0.004, urea \textit{P}=0.04) compared to manual CT (eGFR \textit{P}=0.002, creatinine \textit{P}=0.01, urea \textit{P}=0.03). 
\absdiv{Conclusions}
Our automated approach has greater precision than human-performed manual measurements, which may explain stronger associations with renal function compared to manual measurements. To improve detection of meaningful associations with clinical endpoints in longitudinal studies of OCT, reducing measurement error should be a priority, and automated measurements help achieve this.
\absdiv{Translational relevance}
We introduce a novel choroid segmentation algorithm which can replace manual grading for studying the choroid in renal disease, and other clinical conditions.

\end{abstract}

\begin{document}

\maketitle

\section{Introduction}\label{sec:intro}

Chronic kidney disease (CKD) is a major cause of morbidity and mortality, affecting over 800 million people worldwide \cite{kovesdy2022epidemiology}. CKD is a gradual loss of kidney function, resulting in a failure to filter blood effectively and is associated with alterations in microvascular structure and function \cite{houben2017assessing}. Assessment of microvascular function is therefore helpful in the diagnosis, prognosis and treatment of renal disease. Current clinical evaluation is based on blood and urine testing and invasive kidney biopsy.

The eye and the kidney show a striking resemblance in anatomy, physiology and response to disease \cite{wong2014kidney}, and in particular the choroidal microcirculation reflects the renal microcirculation \cite{farrah2020eye}. For example, they have an analogous vascular endothelium with similarly sized fenestrated vessel walls allowing subretinal fluid exchange in the choroid and blood filtration in the kidneys. Moreover, the circulatory systems in the choroid and renal cortex have similar proportions of blood flow in contrast to their retinal and renal medullary counterparts \cite{nickla2010multifunctional}.

Choroidal thickness (CT) correlates strongly with renal dysfunction in patients with CKD compared with sex- and age-matched healthy and hypertensive individuals \cite{balmforth2016chorioretinal}. This suggests that choroidal thinning may relate to systemic microvascular injury in general and renal injury in particular. An overactive sympathetic drive may contribute to disease progression in CKD \cite{kaur2017sympathetic}, and could explain choroidal thinning since choroidal perfusion is influenced by its autonomic supply \cite{balmforth2016chorioretinal}. These observations suggest the potential clinical utility of choroidal biomarkers in renal disease. There is already a growing interest in choroidal biomarkers derived from optical coherence tomography (OCT) in non-ocular pathology such as sepsis \cite{courtie2020retinal, erikson2017retinal}, diabetes and CKD \cite{balmforth2016chorioretinal, ishibazawa2015choroidal, nakano2020choroid}, and neurodegenerative disease \cite{di2019optical, garcia2019changes, bulut2016choroidal}. 

chorioretinal structures with micro-meter resolution.
Standard OCT imaging focuses on imaging the retina while enhanced depth imaging OCT (EDI-OCT) provides a deeper and stronger signal, visualising the choroid-scleral (C-S) junction with micron resolution. Measuring the choroid manually from OCT images is a time consuming and subjective process. Automated image analysis to extract choroidal measures as potential biomarkers could have substantial clinical utility if it is sufficiently reliable, robust, and reproducible. Consequently, there have been many attempts to validate automated algorithms for quantifying the choroid \cite{mishra2020automated, zhang2015validity, masood2019automatic, zahavi2021evaluation}. However, only a handful of these proposed algorithms demonstrate their clinical utility in disease settings \cite{chen2022application, dolz2016automated}. To the best of our knowledge, such automated algorithms have rarely been applied to longitudinal image sets. This is important as accurate assessment of change over time is essential for tracking disease progression. 

We have previously developed an automated approach to choroid region segmentation \cite{burke2021edge} and here we present our evaluation of its performance and clinical utility in a longitudinal cohort of individuals with end-stage CKD who underwent renal transplant and healthy donors who underwent unilateral nephrectomy. Our primary objective was to compare our automated CT measurements with the current gold standard of manual measurement of CT, as quantitative validation across longitudinal data is an ideal approach to evaluate novel image processing methodologies in the medical domain. 

As a secondary objective, we validate our segmentation algorithm clinically through estimating associations between the choroid and clinical biomarkers of renal function in both cohorts. In order to limit processing time, the choroid is typically measured manually using only thickness. Our approach permits automatic segmentation and subsequent calculation of choroidal thickness and area (CA), so we provide two metrics of the choroid to estimate associations with.

\section{Methods}
\label{sec:materials_methods}

\subsection{Study Population}
We prospectively analysed a longitudinal cohort of healthy kidney donors and patients with end-stage kidney disease undergoing living donor kidney transplantation (NCT0213274) \cite{dhaun2014optical}. Data collection and subsequent analyses were conducted after ethical approval from the South East Scotland Research Ethics Committee, in accordance with the principles of the Declaration of Helsinki and all participants gave informed consent to recruitment. Eligibility criteria for recruitment were (1) donors must be living and healthy throughout the period of analysis, (2) recipients with end-stage CKD have a functional kidney transplant and (3) must be aged 18 or over. To prevent ocular issues confounding our results, our exclusion criteria were (1) any ocular pathology pre-transplant, (2) any previous eye surgery, (3) a refractive error exceeding ±6 dioptres or (4) a diagnosis of diabetes mellitus. We judged image quality by the OSCAR-IB criteria \cite{tewarie2012oscar} and excluded images with B-scan signal quality $\leq$ 15, indistinguishable C-S junction due to speckle noise, or partial image cropping of choroid.

Supplementary figure \ref{fig:datasetAB_flow} shows a flowchart on how the population was selected for this study. At the time of data extraction, 22 donors and 23 recipients were eligible for analysis, after excluding a total of 75 recipients and 5 donors. The majority of exclusions were because there were no eye scans performed for these individuals. This was because, once recruited, they were scheduled for transplantation and then immediately returned to their referring institution after transplant away from Edinburgh, thus making it impractical to retain in the study and re-scan. Each participant was assessed on the day of transplant and then attempts were made at 7 other time periods over the following year (8 time points in total). 

Donors and recipients were organised into two datasets for \textit{performance} evaluation and \textit{clinical} evaluation. We used the entire analysis cohort for performance evaluation of the automated approach against manual grading, resulting in 483 CT measurements for comparison. After excluding 2 donors and 7 recipients due to missing clinical measurements/eye scans post-transplant, or poor quality EDI-OCT images, we used 20 donors and 16 recipients for clinical evaluation, i.e. association estimation with clinical variables related to renal function. Clinical data at each time point included standard biomarkers of renal function, including high sensitivity C-reactive protein (hsCRP), serum urea and creatinine, estimated glomerular filtration rate (eGFR) and urine protein to creatinine ratio. Table \ref{tab:pop_stats_A} shows the population statistics for the entire analysis cohort.

\begin{table*}[!t]\footnotesize
\centering
\begin{tabular}{p{3.1cm}|p{1.55cm}p{1.45cm}p{1.45cm}p{1.45cm}p{1.45cm}p{1.45cm}p{1.45cm}p{1.45cm}}
 & Baseline & 1 Week & 2 Weeks & 4 Weeks & 8 Weeks & 18 Weeks & 28 Weeks & 52 Weeks \\
\hline
\multicolumn{1}{c|}{Demographic}  &  &  &  &  &  &  &  \\
 \underline{Donors}  &  &  &  &  &  &  &  &  \\ 
Sample, n (\%) & 22 (100) & 11 (50) & 2 (9) & 8 (36) & 10 (45) & 8 (36) & 10 (45) & 11 (50) \\
Age, years & 50 ± 11 & - & - & - & - & - & - & - \\
Male sex, n (\%) & 9 (41) & - & - & - & - & - & - & - \\
Approx. Refractive error (D) & 0.49 ± 1.17 & - & - & - & - & - & - & - \\
Daytime (Hr:Min) & 13:50 ± 1:41 & 13:48 ± 1:20 & 12:04 ± 1:34 & 12:48 ± 3:47 & 13:43 ± 2:12 & 13:54 ± 2:00 & 13:30 ± 2:03 & 13:48 ± 2:07 \\
Time from baseline (weeks) & - & 1 ± 0 & 2 ± 0 & 4 ± 1 & 8 ± 2 & 16 ± 4 & 29 ± 3 & 53 ± 6 \\
&  &  &  &  &  &  &  &  \\
 \underline{Recipients}  &  &  &  &  &  &  &  &  \\
Sample, n (\%) & 23 (100) & 13 (61) & 5 (48) & 5 (48) & 10 (43) & 6 (26) & 8 (35) & 8 (35) \\
Age, years & 47 ± 12 & - & - & - & - & - & - & - \\
Male sex, n (\%) & 15 (65) & - & - & - & - & - & - & - \\
Approx. Refractive error (D) & -0.33 ± 1.73 & - & - & - & - & - & - & - \\
Daytime (Hr:Min) & 13:59 ± 2:19 & 12:44 ± 1:42 & 11:45 ± 1:50 & 12:23 ± 2:01 & 11:31 ± 1:13 & 12:29 ± 1:55 & 11:47 ± 1:39 & 12:25 ± 1:31 \\
Time from baseline (weeks) & - & 1 ± 0 & 2 ± 0 & 4 ± 1 & 8 ± 2 & 17 ± 4 & 29 ± 4 & 52 ± 4 \\
\hline
\multicolumn{1}{c|}{Clinical}  &  &  &  &  &  &  &  \\
 \underline{Donors} &  &  &  &  &  &  &  \\
BMI, kg/m$^2$ (\%) & 27 ± 3 (95) & - & - & - & - & - & - & - \\
Systolic BP, mmHg (\%) & 134 ± 14 (86) & - & - & - & - & - & - & - \\
Diastolic BP, mmHg (\%) & 86 ± 18 (86) & - & - & - & - & - & - & - \\
MAP (\%) & 102 ± 15 (86) & - & - & - & - & - & - & - \\
hsCRP, mg/l (\%) & 1 ± 2 (59) & 52 ± 30 (73) & 2 ± 1 (100) & 23 ± 34 (63) & 1 ± 1 (70) & 5 ± 9 (75) & 1 ± 1 (80) & 2 ± 4 (55) \\
Serum urea, mmol/l (\%) & 5 ± 1 (95) & 6 ± 3 (91) & 6 ± 1 (100) & 6 ± 1 (88) & 6 ± 1 (100) & 7 ± 2 (75) & 6 ± 1 (90) & 6 ± 1 (64) \\
Creatinine, $\mu$mol/l (\%) & 69 ± 8 (95) & 108 ± 25 (91) & 102 ± 3 (100) & 89 ± 7 (88) & 102 ± 24 (100) & 101 ± 25 (75) & 100 ± 22 (90) & 91 ± 9 (64) \\
eGFR, ml/min/1.73m$^2$ (\%) & 97 ± 11 (95) & 61 ± 10 (91) & 64 ± 9 (100) & 71 ± 10 (88) & 68 ± 18 (100) & 67 ± 14 (75) & 66 ± 12 (90) & 67 ± 9 (64) \\
Urine P:Cr, mg/mmol (\%) & 3 ± 5 (41) & 185 ± 303 (55) & 1 ± 0 (50) & 3 ± 3 (38) & 1 ± 2 (60) & 1 ± 2 (63) & 1 ± 1 (60) & 1 ± 1 (27) \\
 &  &  &  &  &  &  &  &  \\
 \underline{Recipients} &  &  &  &  &  &  &  \\
BMI, kg/m$^2$ (\%) & 27 ± 4 (87) & - & - & - & - & - & - & - \\
Systolic BP, mmHg (\%) & 137 ± 14 (83) & - & - & - & - & - & - & - \\
Diastolic BP, mmHg (\%) & 82 ± 8 (83) & - & - & - & - & - & - & - \\
MAP (\%) & 101 ± 9 (83) & - & - & - & - & - & - & - \\
hsCRP, mg/l (\%) & 3 ± 3 (48) & 17 ± 20 (38) & 3 ± 3 (40) & 7 ± 10 (100) & 3 ± 2 (30) & 2 ± 2 (50) & 1 ± 1 (38) & 25 ± 52 (75) \\
Serum urea, mmol/l (\%) & 17 ± 8 (87) & 7 ± 2 (77) & 7 ± 2 (100) & 6 ± 1 (100) & 7 ± 2 (90) & 7 ± 2 (83) & 8 ± 2 (88) & 6 ± 1 (88) \\
Creatinine, $\mu$mol/l (\%) & 581 ± 271 (87) & 117 ± 35 (77) & 92 ± 24 (100) & 104 ± 25 (100) & 116 ± 27 (90) & 110 ± 32 (100) & 115 ± 40 (88) & 109 ± 27 (88) \\
eGFR, ml/min/1.73m$^2$ (\%) & 19 ± 28 (87) & 61 ± 15 (77) & 79 ± 16 (100) & 71 ± 14 (100) & 67 ± 20 (90) & 70 ± 21 (100) & 64 ± 23 (88) & 69 ± 13 (88) \\
Urine P:Cr, mg/mmol (\%) & 295 ± 360 (48) & 257 ± 28 (23) & 84  ± 72 (40) & 89 ± 57 (60) & 93 ± 126 (50) & 27 ± 14 (100) & 34 ± 20 (63) & 9 ± 13 (38)
\end{tabular}
\caption{Population statistics for all donors and recipients, rounded to the nearest integer. Where appropriate, values are shown as mean ± SD, and values in parentheses represent proportion of data completeness relative to baseline populations. SD, standard deviation; D, dioptres; Hr:min, hour:minutes; BMI, body mass index; BP, blood pressure; MAP, mean arterial pressure, hsCRP, high sensitivity C-reactive protein; eGFR, estimated glomerular filtration rate; Urine P:Cr, urine protein to creatinine ratio.}
\label{tab:pop_stats_A}
\end{table*}

\subsection{Image Acquisition}

We imaged each participant's right eye using the Heidelberg spectral domain OCT Spectralis Standard Module. Patients were examined between 9am and 5pm and were endeavoured to be followed up at around the same time of day at each time point. The average (and standard deviation) daytime which scans were taken at each time point are listed in table \ref{tab:pop_stats_A}. During each patient's examination, a horizontal line, EDI-OCT B-scan centred at the foveal pit was taken. Each B-scan covered a $30^{\circ}$ (8.7mm) region and was extracted as a $768\times768$ (pixel height$\times$width) high-resolution image for downstream image processing. 

Each scan was captured using active eye tracking with Automatic Real Time (ART) software, averaging up to a maximum of 100 scans to generate a single high-resolution image. EDI-OCT images include the B-scan, alongside an \textit{en face} view of the fundus with a horizontal line representing the B-scan's location on the macula. Although refractive error and axial length were not collected for this population, the built-in scan focus parameter (an approximate value for refractive error) was extracted from the acquisition metadata for each scan and was used for downstream clinical evaluation, and is listed in table \ref{tab:pop_stats_A}. Figure \ref{fig:example_oct_measure} shows an example EDI-OCT scan with results from manual and automated assessment. At each time point, only one EDI-OCT B-scan of each individual's right eye was selected for analysis, and no repeated scans were taken during the same examination.

\subsection{Manual Measurement of Choroidal Thickness}

Manual grading for measuring CT was performed by a single, trained operator across all image sessions for each participant. Manual grading of the choroid is described in the original study carried out by Balmforth, et al. \cite{balmforth2016chorioretinal} Briefly, this was achieved using the calliper tool in HeyEx software (version 1.10.4.0, Heidelberg Engineering, Heidelberg, Germany). Choroid thickness was measured at 3 locations across the macula --- at the fovea and 2000 microns nasal and temporal to the fovea. The choroid region in each B-scan was defined as the area between the lower surface of the junction between the hyperreflective retinal pigment epithelium (RPE) layer and Bruch's membrane complex (RPE-C junction), and the upper surface of the junction between the choroid and sclera (C-S junction).

\begin{figure*}[!tb]
    \centering
    \includegraphics[width = 0.8\textwidth]{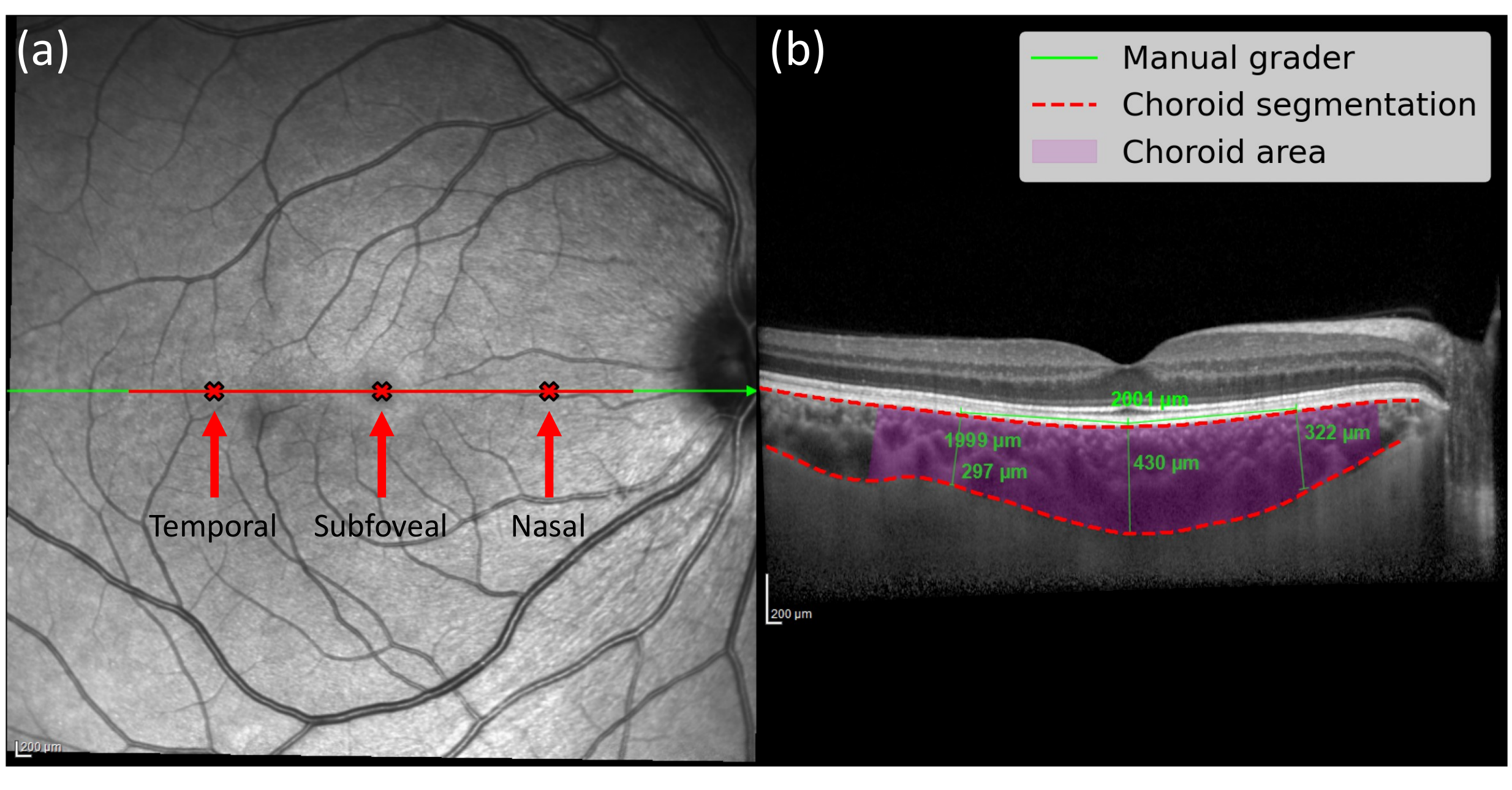}
    \caption{EDI-OCT scan of a donor's right eye at baseline. (a) \textit{En face} view of the fundus with the location of the B-scan in green and markers for measuring CT and CA in red. (b) B-scan showing chorioretinal structures with manual and automated assessment shown.}
    \label{fig:example_oct_measure}
\end{figure*}

Manual grading used the calliper tools on the software to mark the centre of the fovea at the level of the photoreceptor outer segment and mark 2000 microns temporal and nasal to this point, along the RPE-C junction. Choroid thickness was defined as the straight line distance between the RPE-C and C-S junctions, measured locally \textit{perpendicular} to the RPE-C junction. Figure \ref{fig:example_oct_measure} shows three CT measurements provided by the manual grader in green, including the reference lines measuring approximately 2000 microns temporal and nasal to the foveal pit along the RPE-C junction. 

\subsection{Automated Assessment of Choroidal Thickness}

\begin{figure*}[!tb]
\centering
\includegraphics[width=\textwidth]{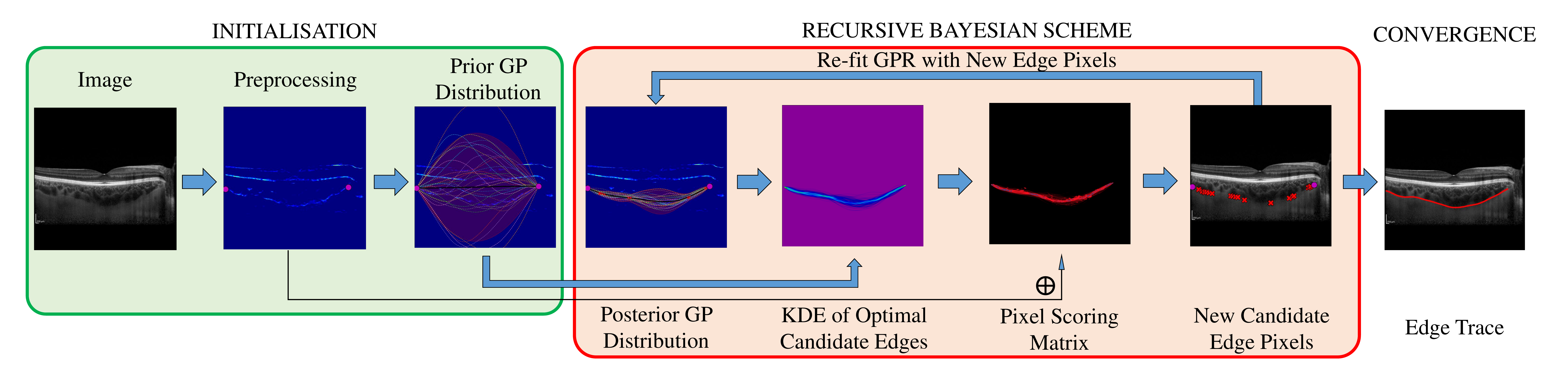}
\caption{Schematic diagram of the automated approach segmenting the C-S junction of the choroid on an EDI-OCT B-scan \cite{burke2021edge}. GP, Gaussian process; GPR, Gaussian process regressor; KDE, kernel density estimate.}
\label{fig:GPET_schematic_diag}
\end{figure*}

Our automated choroid segmentation algorithm performed individual edge tracing to obtain the upper and lower choroid boundaries for each B-scan. The algorithm modelled each boundary using Gaussian process regression, building a posterior predictive function through a recursive Bayesian scheme and image gradient scoring mechanism to detect edge pixels \cite{burke2021edge}. Figure \ref{fig:GPET_schematic_diag} presents a schematic diagram of the image analysis pipeline of the automated approach. Preprocessing was done using median filter denoising and contrast enhancement using contrast limited adaptive histogram equalisation. Approximate edge maps were then computed using discrete derivative and morphological erosion and dilation filters.

Automated CT was measured in two different ways, yielding \textit{perpendicular} CT measurements and \textit{parallel} CT measurements. \textit{Perpendicular} automated CT measured locally \textit{perpendicular} to the RPE-C junction, while \textit{parallel} automated CT measured parallel to the manual measurements to mimic any potential angular error made by the manual grader. See for example the nasal measurement in figure \ref{fig:major_disc_examples}(c), where perpendicular automated CT is shown in red, parallel automated CT in blue, and manual CT in green. We measured CT in the same subfoveal, temporal and nasal locations as manual grading. We measured CA by computing the area of pixels within a 3000 micron radius centred at the foveal pit, in accordance with the Early Treatment Diabetic Retinopathy Study area of $6000 \times 6000$ microns \cite{early1991early}. 

\subsection{Statistical Analysis}

\subsubsection{Performance evaluation}
 We measured agreement between approaches by comparing (parallel) automated CT to manual CT and performed Passing-Bablok analysis \cite{passing1983new} to determine any systematic or proportional differences between the two approaches. We also investigated residuals, calculated by subtracting manual CT from automated CT (automated - manual), using a Bland-Altman plot \cite{bland1986statistical}. Mean absolute error (MAE), Pearson and intra-class correlations were also computed between both approaches, stratified by cohort and macular location. To compare the consistency of measurements across longitudinal data, we investigated the angular deviation made by the manual grader. This was done by computing angles between the calliper measurements made from manual grading against the perpendicular automated CT measurements. Automated assessment of the choroid and quantitative analyses comparing manual and automated measurements were done in Python (version 3.11.3).

Rahman \cite{rahman2011repeatability} found an unsigned residual of $32\mu$m as a threshold suggested to exceed inter-observer variability between graders of EDI-OCT images. Therefore, we defined major discrepancies between automated and manual CT to be any residual greater than $32\mu$m in absolute value. These were selected for external adjudication by a clinical ophthalmologist I.M. For each discrepancy, I.M. was shown two CT measurements displayed on the corresponding choroid, and was blinded to which approach was used in each case. He was asked to rate each measurement, and the overall visibility of the C-S junction, using an ordinal scale: ``bad'', ``okay'' and ``good''. I.M. was also asked to select the approach which was preferred --- options for both or neither approaches were also available.

\subsubsection{Clinical evaluation}
We investigated how CT and CA measurements changed over time in donors and recipients, and estimated linear associations with clinical variables related to renal function using Pearson correlations --- specifically eGFR, serum creatinine and serum urea. Other clinical measures were too incomplete and inconsistently collected to be analysed with the current sample size. CT was averaged across macular locations, resulting in a single value of thickness for each scan. We examined the difference in CT and CA between measurements at baseline and 1-year post-transplant, testing for difference in their means longitudinally via the Student's t-test for dependent, paired samples. We specified a \textit{P}-value of 0.05 as the threshold for statistical significance.

We also looked for significant linear associations between markers of renal function and the choroid. We assumed a nested structure to the data and used linear mixed-effects (LME) models, setting an individual random effect to model the variation among individuals, accounting for age, sex, approximate refractive error (measured in dioptres) and the time of day the scan was taken, relative to when their baseline scan was taken (measured in decimal hours). For each clinical measure of renal function, we fitted an LME model for each choroidal image measure in turn as one of the associated independent variables. LME modelling was carried out using R (version 4.2.2) \cite{lmer_R}. We identify independent variables as statistically significant predictors if their associated 95\% confidence interval excludes 0.

\begin{table*}[!tb]
\centering
\begin{tabular}{llllllll}
\toprule
\multirow{2}{*}{Metric} & \multirow{2}{*}{All} & \multicolumn{3}{c}{Donors} & \multicolumn{3}{c}{Recipients} \\ \cmidrule(l){3-5}\cmidrule(l){6-8}
& & Temporal & Subfoveal & Nasal & Temporal & Subfoveal & Nasal \\
\midrule
Residual CT ($\mu$m) & 1.8 ± 22.0 & 5.1 ± 18.4 & -5.9 ± 17.4 & 2.1 ± 18.8 &  2.9 ± 24.3 & 3.0 ± 28.1 & 3.8 ± 21.0\\
MAE CT ($\mu$m) & 14.1 ± 16.9 & 13.8 ± 13.2 & 13.4 ± 12.5 & 12.9 ± 13.8 &  13.5 ± 20.4 & 17.3 ± 22.3 & 13.8 ± 16.2\\
Pearson, $r_p$ & 0.97 & 0.96 & 0.98 & 0.97 & 0.92 & 0.94 & 0.96 \\
ICC(3,1) & 0.96 & 0.95 & 0.97 & 0.96 & 0.92 & 0.94 & 0.96\\
\bottomrule
\end{tabular}
\caption{Performance evaluation between automated and manual CT, stratified by cohort and macular location. Where appropriate, values are shown as mean ± SD. All Pearson and intra-class correlations were statistically significant. SD, standard deviation; MAE, mean absolute error; CT, choroid thickness; ICC, intra-class correlation.}
\label{tab:table_perf}
\end{table*}

\section{Results}

\subsection{Performance Evaluation with Choroidal Thickness}

Performance results between automated and manual CT measurements are shown in table \ref{tab:table_perf}. There was good agreement between manual and automated measures of CT, with an average residual of 1.8 ± 22.0$\mu$m and MAE of 14.1 ± 16.9$\mu$m across all 483 CT measurements. Moreover, Pearson and intra-class correlation show excellent correlations between both approaches. The largest residuals were at the subfoveal macular location.

\begin{figure*}[!tb]
     \centering
    \includegraphics[width=\textwidth]{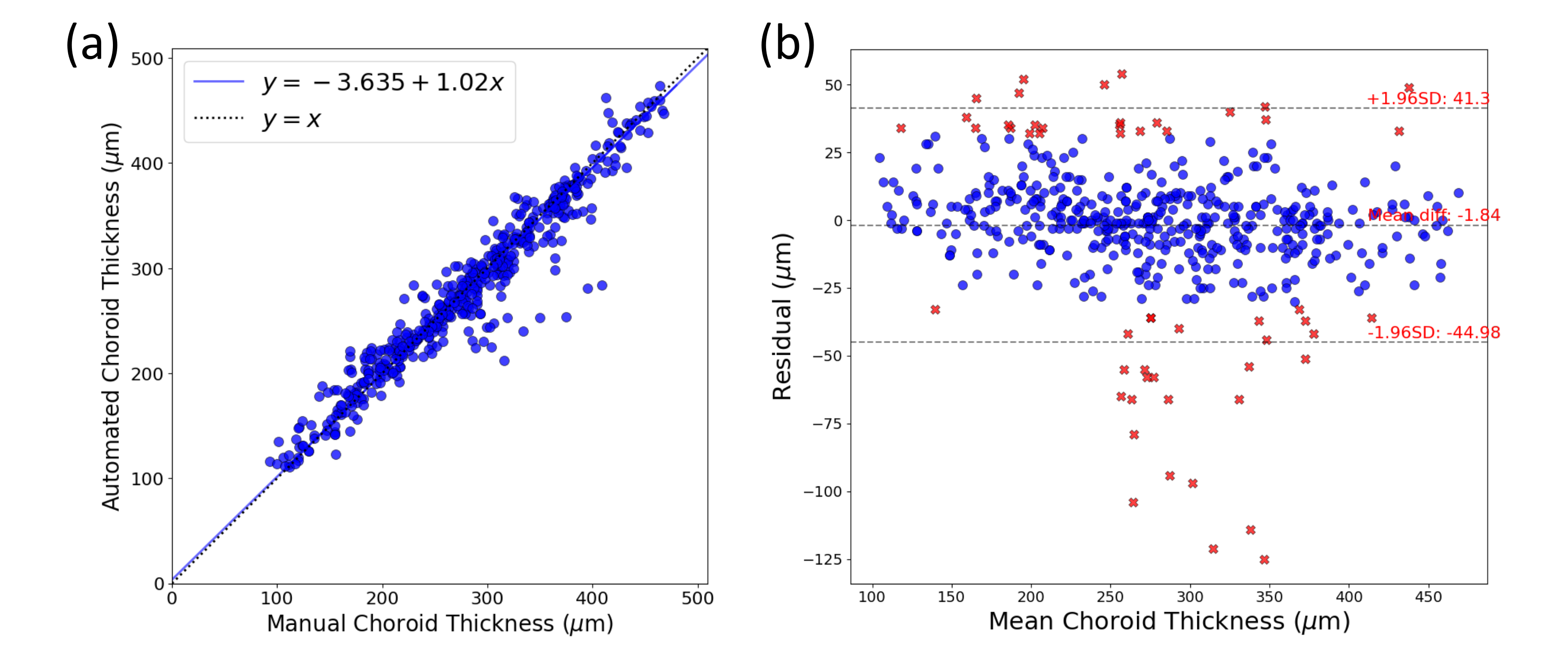}
    \caption{(a) Correlation plot of manual and (parallel) automated CT measurements. (b) Bland-Altman plot showing residual values, with 52 major discrepancies highlighted as red crosses.}
    \label{fig:BA_PB_plots}
\end{figure*}

Passing-Bablok analysis computed a slope value of 1.02 (95\% confidence interval of 0.99 to 1.04) and an intercept of -3.64$\mu$m (95\% confidence interval of -10.78$\mu$m to 3.50$\mu$m) with an $R^2$ value of 0.93 (figure \ref{fig:BA_PB_plots}(a)). These 95\% confidence intervals include 1 and 0, respectively, suggesting that the automated and manual approaches are equivalent \cite{ranganathan2017common, passing1983new}. See supplementary figures \ref{fig:corr_plot_loc} and \ref{fig:corr_plot_cohort} for correlation plots stratified by macular location and cohort. Bland-Altman analysis on figure \ref{fig:BA_PB_plots}(b) computed an average residual of 1.84$\mu$m with limits of agreement of -44.98$\mu$m to 41.30$\mu$m. Only 10.78\% of all 483 CT measurements (52 / 483) exceeded an unsigned residual of $32\mu$m \cite{rahman2011repeatability}.

\begin{table}[!tb]
\centering
\begin{tabular}{@{}lccc@{}}
\toprule
\multicolumn{1}{c}{\multirow{2}{*}{\begin{tabular}[c]{@{}c@{}}C-S Junction \\ Visibility\end{tabular}}} & \multicolumn{2}{c}{Score} \\ \cmidrule(l){2-3}
& Automated & Manual \\ 
\midrule
Good ($N=9$) & 8 & 2 \\
Okay ($N=15$) & 9 & 13 \\
Bad ($N=28$) & 24 & 15 \\
Total ($N=52)$ & 41 & 30 \\
\midrule
\midrule
Method & \multicolumn{2}{c}{Discrepancy Measurement} \\
\midrule
Automated & \multicolumn{2}{l}{Good: 32, Okay: 19, Bad: 1} \\
Manual   & \multicolumn{2}{l}{Good: 24, Okay: 18, Bad: 10} \\
\bottomrule
\end{tabular}%
\caption{(Top) Numerical score of both approaches from masked adjudication of 52 major discrepancies, stratified by the visibility of the C-S junction. (Bottom) Qualitative preference scores for both approaches.}
\label{tab:outlier_anal}
\end{table}

\begin{figure*}[!tb]
     \centering
    \includegraphics[width=0.8\textwidth]{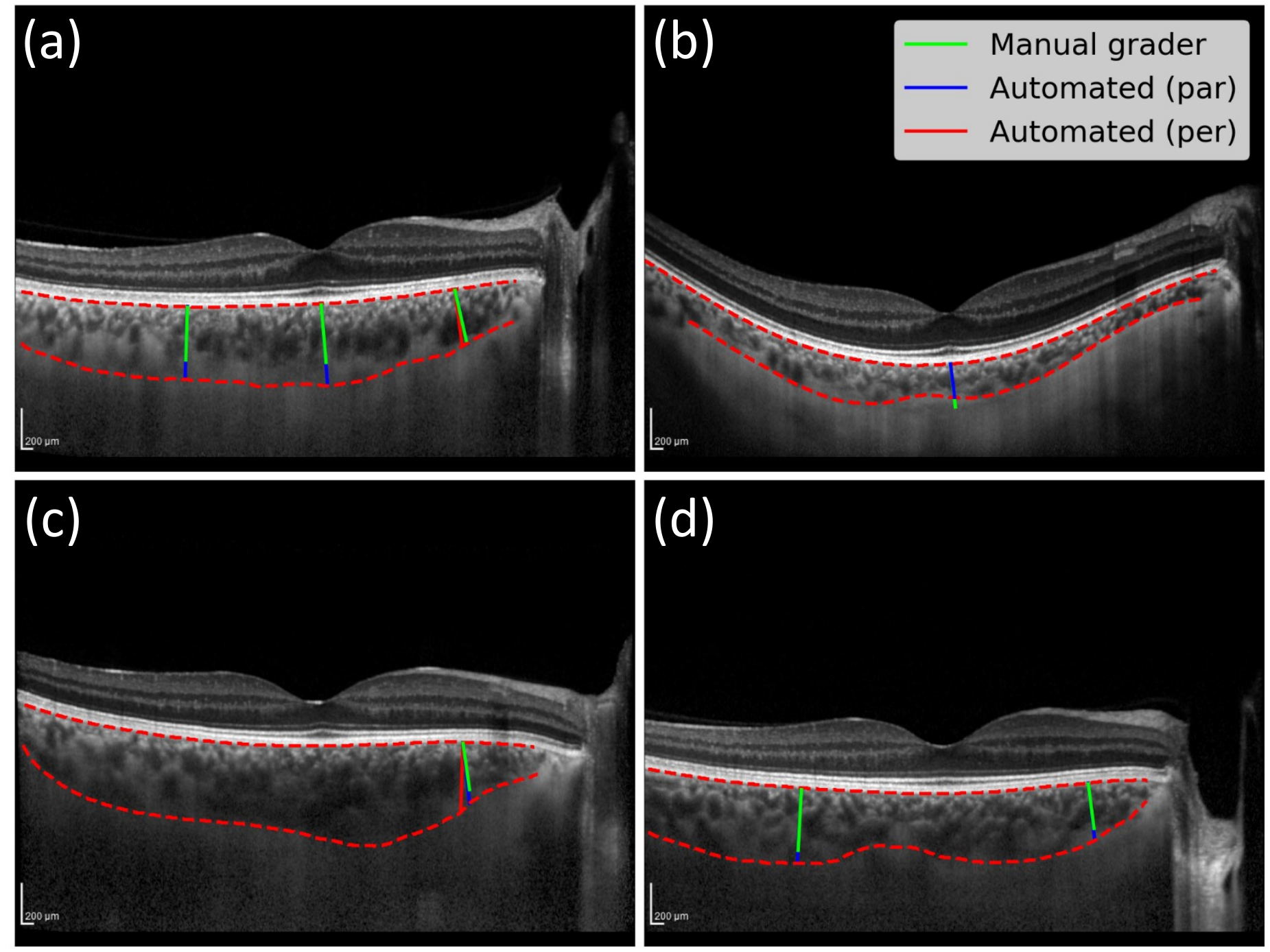}
    \caption{A selection of major discrepancies between automated and manual CT measurements. Red lines represent perpendicular automated CT, blue represent parallel automated CT and green represent manual CT. The lines in red are shown in (a,c) to demonstrate the observed angular deviation between perpendicular automated CT and manual CT. (a) Temporal and subfoveal major discrepancies from a donor, residuals $97\mu$m and $114\mu$m, respectively. Nasal angular deviation (5.4$^\circ$) resulted in perpendicular and parallel residuals as $22\mu$m and $2\mu$m, respectively. (b) Subfoveal major discrepancy from a recipient, residual $-47\mu$m. (c) Nasal angular deviation (8.1$^\circ$) shows perpendicular and parallel residuals as $106\mu$m and $79\mu$m, respectively. (d) Temporal and nasal major discrepancy from a donor, residuals $66\mu$m and $65\mu$m, respectively.}
    \label{fig:major_disc_examples}
\end{figure*}

Table \ref{tab:outlier_anal} presents the results from the external adjudication of all 52 major discrepancies. Overall, the automated CT measurements scored higher in terms of preference, where 79\% (41/52) of the automated thicknesses were preferred compared to 58\% (30/52) for manual thicknesses. Moreover, the automated approach scored higher in terms of quality, with only 1 measurement judged as ``bad'', versus 10 measurements judged ``bad'' for the manual approach. Figure \ref{fig:major_disc_examples} shows four major discrepancies from four different choroids. Figure figure \ref{fig:major_disc_examples}(a) shows the largest observed discrepancies --- temporal ($97\mu$m) and subfoveal ($114\mu$m) --- with adjudicator I.M. preferring the automated measurements.

\begin{figure*}[!tb]
     \centering
    \includegraphics[width=\textwidth]{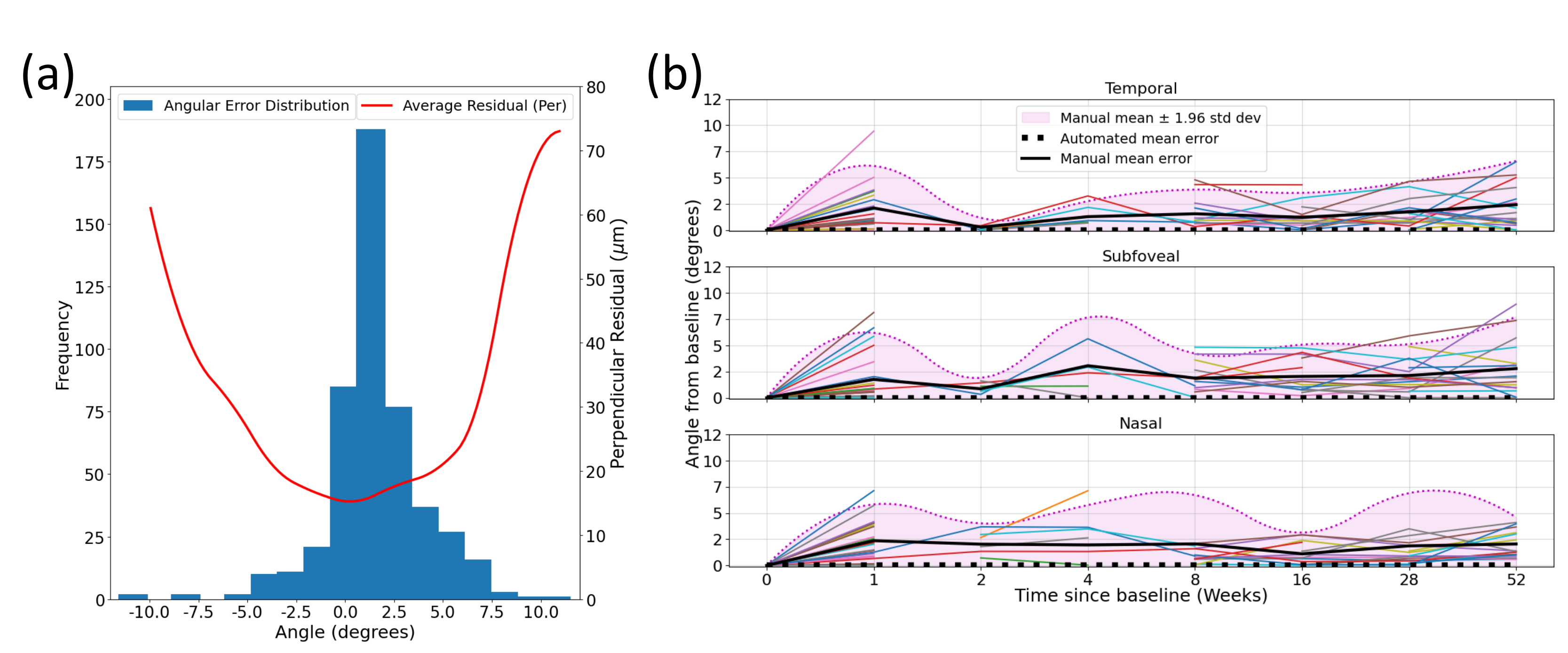}
    \caption{(a) Distribution of angular errors from manual CT measurements (blue). Average unsigned difference between manual and perpendicular automated CT (red). (b) Unsigned, longitudinal within-patient angular deviations, relative to baseline measurements. Coloured lines represent distinct individuals using manual measurements.}
    \label{fig:angle_disc}
\end{figure*}

We found that deviation in the angle of manual measurement from perpendicular led to a disproportionate error between perpendicular automated CT and manual CT. This is illustrated in figure \ref{fig:angle_disc}(a) which shows the distribution (mean 0.5 and standard deviation 2.5) of angular errors made from manual grading. Absolute differences between manual CT and perpendicular automated CT grew quadratically as the angular error deviated from 0. We also found large deviation in manual grading when tracking the longitudinal, within-patient angular difference from baseline (figure \ref{fig:angle_disc}(b)). Distributions at each time point were estimated through empirical mean and standard deviation values. Note that the mean fluctuation in manual grading was always above 0 in comparison to the mean fluctuation from the automated approach which was constant at 0. 

\subsection{Choroidal Association with Renal Function}
\begin{figure}[!tb]
    \centering
    \includegraphics[width = 0.495\textwidth]{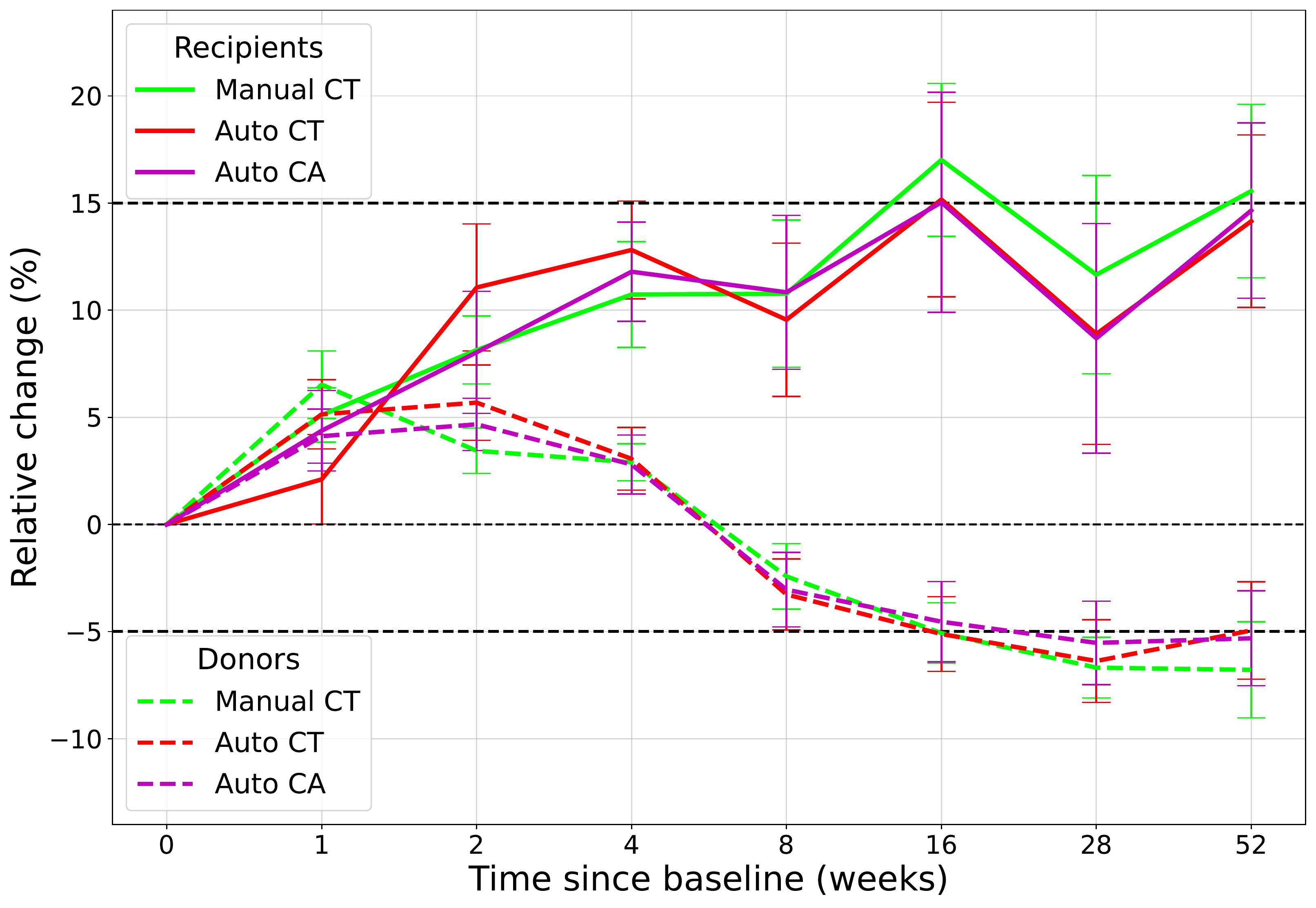}
    \caption{Longitudinal change in the choroid, measured relative to baseline measurements. Donors are shown as dashed lines and recipients as solid lines, with standard errors plotted at each time point.}
    \label{fig:longitudinal_fig}
\end{figure}

All measures of renal function and choroidal measurements were significantly different 1-year post-transplant compared to baseline measurements in both cohorts (table \ref{tab:longitudinal_ttest}). Figure \ref{fig:longitudinal_fig} shows the average percentage change in choroidal measurements from baseline measurements in both cohorts. In transplant recipients, there was a significant increase in choroidal thickness of 12.8 ± 4.8\% (median: 11.4\%) from baseline after 4 weeks post-transplant using automated CT ($\textit{P}$=0.02). After 1 year post-transplant, we observed a significant increase of 14.1 ± 11.4\% (median: 10.7\%) from baseline ($\textit{P}$=0.01). Interestingly, we observed a significant increase of 5.1 ± 5.6\% (median: 5.4\%) 1 week post-transplant in the choroid of donors ($\textit{P}$=0.01). The average choroid remained inflated relative to baseline measurements and at some point between 4 and 8 weeks post-transplant the choroid deflated. After 1 year post-transplant, we observed a significant decrease of -4.9 ± 7.9\% (median: -3.5\%) in automated CT ($\textit{P}$=0.01).

\begin{table*}[!tb]
\centering
\begin{tabular}{lllllll}
\toprule
 & \multicolumn{3}{c}{Donors} & \multicolumn{3}{c}{Recipients} \\ \cmidrule(l){2-4}\cmidrule(l){5-7}
\multirow{1}{*}{Measurement} & \multicolumn{1}{c}{Baseline} & \multicolumn{1}{c}{1 year PT} & \multicolumn{1}{c}{$\textit{P}$-value} & \multicolumn{1}{c}{Baseline} & \multicolumn{1}{c}{1 year PT} & \multicolumn{1}{c}{$\textit{P}$-value} \\ 
\midrule
Automated CA (mm$^2$) & 1.59 $\pm$ 0.41 & 1.49 $\pm$ 0.41 & 0.01 & 1.58 $\pm$ 0.41 & 1.74 $\pm$ 0.46 & 0.01 \\
Automated CT ($\mu$m) & 278 $\pm$ 69 & 259 $\pm$ 68 & 0.01 & 271 $\pm$ 70 & 301 $\pm$ 75 & 0.01 \\
Manual CT ($\mu$m) & 278 $\pm$ 61 & 258 $\pm$ 62 & 0.006 & 262 $\pm$ 64 & 292 $\pm$ 66 & 0.008\\
\midrule
eGFR (ml/min/1.73m$^2$) & 91 $\pm$ 5 & 64 $\pm$ 7 & <0.001 & 8 $\pm$ 3 & 69 $\pm$ 14 & <0.001 \\
Creatinine ($\mu$mol/l) & 69 $\pm$ 9 & 92 $\pm$ 10 & <0.001 & 659 $\pm$ 196 & 109 $\pm$ 29 & <0.001 \\
Urea (mmol/l) & 5.0 $\pm$ 0.8 & 6.8 $\pm$ 1.3 & 0.002 & 18.5 $\pm$ 9.7 & 6.7 $\pm$ 1.7 & 0.005 \\
\bottomrule
\end{tabular}
\caption{Choroidal measurements at baseline and one year post-transplant (PT). $\textit{P}$-values calculated using Student's dependent t-test for related samples.}
\label{tab:longitudinal_ttest}
\end{table*}

In the recipient cohort, all choroidal measurements linearly correlated with eGFR, serum creatinine and serum urea with statistical significance. All Pearson correlation coefficients with renal function were stronger in automated choroid measurements than in manual ones (table \ref{tab:clinical_correlation}). There were poor and insignificant linear correlations found between the choroid and renal function in the donor cohort, as seen in supplementary table \ref{tab:clinic_corr_nonsig}, which also show the corresponding $\textit{P}$-values for all correlation analyses performed.

\begin{table}[!tb]
\centering
\begin{tabular}{llll}
\toprule
 & \multicolumn{3}{c}{Pearson, $r_p$} \\ \cmidrule(l){2-4}
\multicolumn{1}{l}{Image Measure} & eGFR & Creatinine & Urea \\
 \midrule
Automated CA & 0.72 & -0.77 & -0.73 \\
Automated CT & 0.81 & -0.76 & -0.74\\
Manual CT & 0.69 & -0.76 & -0.72 \\
\bottomrule
\end{tabular}
\caption{Pearson correlation coefficients between choroidal measurements and eGFR, serum creatinine and serum urea in the recipient cohort. All correlations presented here were significant such that $\textit{P}$<0.05 apart from eGFR using manual CT. }
\label{tab:clinical_correlation}
\end{table}

Using all choroidal measurements, in transplant recipients LME modelling showed evidence of a statistically significant, positive linear association with eGFR, and negative associations with serum creatinine and urea levels. Table \ref{tab:mixed_lm_recips} shows the model summary's for predicting eGFR, serum creatinine and serum urea using each choroidal measurement in turn. Models using automated measures had greater conditional $R^2$ values --- a measure analogous to the $R^2$ value for linear models but taking into account the random and fixed effects variables --- and their respective independent variables showed stronger association to predicting renal function. Interestingly, the time difference between follow-up and baseline image acquisition was a statistically significant predictor of all markers of renal functions for models using all choroidal metrics.

There were no statistically significant associations found between any choroidal measurement and markers of renal function in the donor cohort. However, sex was associated with predicting serum creatinine using automated choroidal measurements, with weak statistical significance. These LME results can be found in supplementary table \ref{tab:mixed_lm_donors}.

\begin{table*}[!tb]
\centering
\begin{tabular}{@{}llllllllll@{}}
\toprule
\multicolumn{1}{c}{Response variable $\rightarrow$} &
  \multicolumn{3}{c}{eGFR} &
  \multicolumn{3}{c}{Creatinine} &
  \multicolumn{3}{c}{Urea} \\ \cmidrule(l){1-1}\cmidrule(l){2-4}\cmidrule(l){5-7}\cmidrule(l){8-10} 
\multicolumn{1}{c}{Independent variable $\downarrow$} &
  \multicolumn{1}{c}{$\hat{\beta}$} &
  \multicolumn{1}{c}{95\% CI} &
  \multicolumn{1}{c}{$P$-value} &
  \multicolumn{1}{c}{$\hat{\beta}$} &
  \multicolumn{1}{c}{95\% CI} &
  \multicolumn{1}{c}{$P$-value} &
  \multicolumn{1}{c}{$\hat{\beta}$} &
  \multicolumn{1}{c}{95\% CI} &
  \multicolumn{1}{c}{$P$-value}\\
  \midrule

Intercept & -0.10 & (-1.38, 1.17) & 0.84 & 0.09 & (-1.18, 1.36) & 0.86 & 0.03 & (-0.49, 0.55) & 0.90 \\
Age & 0.98 & (-0.47, 2.43) & 0.15 & -1.11 & (-2.54, 0.31) & 0.10  & -0.30 & (-0.95, 0.34) & 0.29 \\
Sex (Male) & -0.24 & (-1.55, 1.07) & 0.66 & 0.27 & (-1.03, 1.57) & 0.60  & 0.22 & (-0.31, 0.76) & 0.34 \\
\textbf{Daytime from Baseline} & -0.40 & (-0.59, -0.21) & \underline{\textbf{<0.001}} & 0.59 & (0.39, 0.80) & \underline{\textbf{<0.001}}  & 0.57 & (0.32, 0.81) & \underline{\textbf{<0.001}} \\ 
Approx. Refraction & -0.86 & (-2.17, 0.46) & 0.114 & 0.92 & (-0.37, 2.21) & 0.12  & 0.31 & (-0.28, 0.89) & 0.24 \\
\textbf{Automated CA} & 1.54 & (0.97, 2.10) & \underline{\textbf{<0.001}} & -1.52 & (-2.12, -0.92) & \underline{\textbf{<0.001}}  & -0.60 & (-1.09, -0.11) & \underline{\textbf{0.02}} \\

Patient Std. Dev. & 1.63 & \multicolumn{2}{l}{Cond. $R^2$: 0.93} & 0.60 & \multicolumn{2}{l}{Cond. $R^2$: 0.91} & 0.55 & \multicolumn{2}{l}{Cond. $R^2$: 0.57} \\
\midrule

Intercept & -0.10 & (-1.19, 0.99) & 0.83 & 0.07 & (-0.91, 1.05) & 0.84  & 0.03 & (-0.48, 0.53) & 0.90 \\
Age & 0.74 & (-0.51, 1.99) & 0.20 & -0.80 & (-1.92, 0.31) & 0.12 & -0.27 & (-0.90, 0.36) & 0.33 \\
Sex (Male) & -0.31 & (-1.43, 0.81) & 0.49 & 0.31 & (-0.69, 1.30) & 0.44  & 0.28 & (-0.25, 0.81) & 0.24 \\
\textbf{Daytime from Baseline} & -0.46 & (-0.67, -0.24) & \underline{\textbf{<0.001}} & 0.65 & (0.42, 0.88) & \underline{\textbf{<0.001}} & 0.55 & (0.30, 0.80) & \underline{\textbf{<0.001}} \\ 
Approx. Refraction & -0.43 & (-1.55, 0.68) & 0.37 & 0.46 & (-0.54, 1.46) & 0.27  & 0.18 & (-0.35, 0.71) & 0.42 \\
\textbf{Automated CT} & 1.15 & (0.51, 1.78) & \underline{\textbf{<0.001}} & -1.01 & (-1.62, -0.40) & \underline{\textbf{0.004}} & -0.54 & (-1.06, -0.02) & \underline{\textbf{0.04}} \\

Patient Std. Dev. & 1.37 & \multicolumn{2}{l}{Cond. $R^2$: 0.88} & 1.05 & \multicolumn{2}{l}{Cond. $R^2$: 0.81} & 0.50 & \multicolumn{2}{l}{Cond. $R^2$: 0.52} \\

\midrule
Intercept & -0.10 & (-1.22, 1.02) & 0.83 & 0.07 & (-0.89, 1.03) & 0.84 & 0.03 & (-0.47, 0.52) & 0.90 \\
Age & 0.70 & (-0.59, 1.98) & 0.23 & -0.72 & (-1.82, 0.38) & 0.14 & -0.27 & (-0.89, 0.35) & 0.33 \\
Sex (Male) & -0.29 & (-1.44, 0.86) & 0.55 & 0.27 & (-0.71, 1.24) & 0.48  & 0.28 & (-0.24, 0.80) & 0.23 \\ 
\textbf{Daytime from Baseline} & -0.47 & (-0.68, -0.25) & \underline{\textbf{<0.001}} & 0.66 & (0.43, 0.90) & \underline{\textbf{<0.001}} & 0.56 & (0.31, 0.80) & \underline{\textbf{<0.001}} \\ 
Approx. Refraction & -0.41 & (-1.56, 0.74) & 0.41 & 0.41 & (-0.57, 1.39) & 0.29 & 0.18 & (-0.35, 0.70) & 0.42 \\
\textbf{Manual CT} & 1.11 & (0.45, 1.76) & \underline{\textbf{0.002}} & -0.89 & (-1.54, -0.24) & \underline{\textbf{0.01}} & -0.56 & (-1.06, -0.05) & \underline{\textbf{0.03}} \\

Patient Std. Dev. & 1.39 & \multicolumn{2}{l}{Cond. $R^2$: 0.87} & 0.99 & \multicolumn{2}{l}{Cond. $R^2$: 0.79} & 0.49 & \multicolumn{2}{l}{Cond. $R^2$: 0.52} \\ 

\bottomrule
\end{tabular}%
\caption{LME summary tables for predicting eGFR, serum creatinine and serum urea in transplant recipients using each choroidal measurement. Results for the donor cohort are shown in supplementary table \ref{tab:mixed_lm_donors}, but were omitted due to a lack of significant results with the choroid. $\hat{\beta}$ represent standardised model coefficients, with their corresponding 95\% confidence intervals and associated $\textit{P}$-values. Statistically significant variables are shown in bold and underlined, as well as their corresponding $\textit{P}$-values.}
\label{tab:mixed_lm_recips}
\end{table*}

\section{Discussion}

We found that automated CT measurements agreed well with manual ones in general, but have greater precision --- especially when applied to a longitudinal series of images. Much of the superior precision from automated CT results from eliminating inconsistent measurement angle, since the automated method is always able to measure CT locally perpendicular at the exact same location across all time points for each patient --- this is an important attribute for measurement protocol so that we can account for choroidal curvature, or if the choroid does not appear predominantly horizontal in the image. In contrast, error in the angle of manual measurement had a disproportionate impact. 
 
The RPE-C junction is almost never parallel to the C-S junction, making it a challenge for humans to measure perpendicular angles by eye across repeated measures, which is a major source of error. Furthermore, the different axial and lateral resolutions in each B-scan can result in the slightest change in marking a lateral position during manual grading to misrepresent the true CT. Figures \ref{fig:major_disc_examples}(a,c) represent two exemplar cases, where the residuals between manual CT and perpendicular/parallel automated CT measurements are very different.

Importantly, the superior precision of the automated method translated into stronger associations with renal function, at least in the recipient cohort where statistical significance was observed. This illustrates the necessity for reliable and reproducible measurements in longitudinal studies and their subsequent impact on predicting clinical outcomes. 

Much of the discrepancy between manual and automated CT arose from choroidal features that are inherently difficult to measure. For example, the C-S junctions of larger choroids (figure \ref{fig:major_disc_examples}(a, c, d)) are prone to poor quality image acquisition due to higher incidence of poor signal penetration. However, the external adjudicator preferred the automated measurements over the manual grader for these major discrepancies. Furthermore, the C-S junction may be obscured in choroids whose posterior of Haller's layer is low contrast (figure \ref{fig:major_disc_examples}(b)) --- here, the external adjudicator preferred the manual grader's measurement. The choroid in figure \ref{fig:major_disc_examples}(a) represents a source of major disagreement between the manual grader and the automated approach. The manual grader has defined the C-S junction as the boundary below the most visible posterior vessels (green), while the automated approach has successfully identified vessels with much lower visibility further below the clearer vessels (red), of which the external adjudicator agreed with.

The choroids in figure \ref{fig:major_disc_examples} represent just over half (51.9\%) of all major discrepancies lying outside the threshold suggested to exceed inter-observer variability between graders of EDI-OCT images \cite{rahman2011repeatability}. That is, most outliers came from the same four choroid at different time points. Moreover, of the 28 thickness measurements from choroids described as having poor visibility, 20 of them came from the larger choroids (figure \ref{fig:major_disc_examples}(a, c, d)).

From both correlation analysis and LME modelling, automated choroidal measurements were found to correspond with markers of renal function over time better than manual CT, probably because of reduced measurement error. All choroidal measurements changed substantially over time for all study participants, as did eGFR, serum creatinine and serum urea. This is consistent with the choroid reflecting renal function during treatment of CKD. The choroid, and indeed CT/CA, does vary naturally during the day due to diurnal variation. Tan et al. \cite{tan2012diurnal} describe the average change in CT across daytime hours to be approximately 8.5 ± 5.2\%. The change in automated CT in transplant recipients 1 year post-transplant was 14.1 ± 11.4\% (14.7 ± 11.2\% for CA). This suggests that choroidal inflation in transplant recipients cannot be explained fully through diurnal variation, but potentially through improved renal function \cite{choi2020strong}. However, the same statement for healthy kidney donors cannot be made as confidently, where we only observed an average decrease in CT of -4.9 ± 7.9\% (-5.3 ± 7.7\% for CA).

Although, we must note that there are big fluid shifts which occur at the time of transplantation, and we have not been able to account for any change in body fluid volume post-transplant and what impact that may have had on the choroid. However, these fluid shifts would only impact the choroid in the short term, and our follow-up data is of sufficient time course to exclude any potential long term effect from body fluid volume on the significant changes we see in the choroid.

In each LME model, we accounted for age, sex, approximate refractive error, and the intra-participant relative-to-baseline daytime each scan was taken (relative daytime). The latter two measurements were used to assess the impact of myopia and diurnal variation in both cohorts for predicting renal function using the choroid. Relative daytime proved to be a significant predictor variable for markers of renal function alongside all choroidal measurements, while approximate refractive error did not. Note however that the standardised model coefficients are almost always larger in absolute value for the choroidal measurements versus relative daytime.

In every model there was an opposite effect of relative daytime compared with the choroid, suggesting that the size of the choroid is negatively offset such that the CT and CA in a follow-up scan taken later in the day (relative to the baseline acquisition time) is decreased in absolute value --- and vice versa for follow-up scans taken earlier than the original baseline scan. This makes sense given we know that the choroid decreases marginally through the course of the day \cite{tan2012diurnal}. 

This suggests that while choroidal measurements still play a significant role in predicting markers of renal function for transplant recipients, it is important that daytime be recorded for downstream analysis, and image acquisition try be performed around the same time of day so as not to diminish any longitudinal signal present in the data. Note however that removing relative daytime still resulted in the choroid as a statistically significant predictor variable but the overall model fit was marginally worse (data not shown). 

It may be that the immediate inflation of the choroid in the first 4 weeks after kidney donation in donors could be a result of a prompt systemic, cardiovascular response to the unilateral nephrectomy. While from 8 weeks post-transplant we observed choroidal thinning from all measurements, consistent with previous literature \cite{mule2019association}, it is within the first month in a donor's recovery period that compensatory renal hypertrophy takes place \cite{anderson1991short, rojas2019compensatory}. These observations align with Choi and Kim \cite{choi2020strong}, who cautiously suggest that a decrease in subfoveal CT is associated with a decline in renal function but its increase is associated with renal hypertrophy, albeit in patients with diabetic retinopathy. 
 
Low signal quality, vessel shadowing, lack of signal penetration and the potential for extravascular fluid to obscure and occlude the boundary all contribute to the challenging task of tracing the C-S junction in OCT images. However, the automated segmentation methodology has the ability to overcome many of these problems due its tunable hyperparameters and model uncertainty quantification \cite{burke2021edge}. For example, the Bayesian recursive scheme allows uncertainty in obscured or occluded regions to be quantified and correctly interpolated. This feature is advantageous for choroid region segmentation. Moreover, the methodology provides an analytical, functional form for the edge which is explainable, interpretable and reproducible --- features which tend to be absent from many deep learning algorithms in medical image segmentation \cite{renard2020variability}.

There were a number of limitations in this study. Firstly, our two cohorts were limited in sample size and not all participants or clinical variables were collected at every time point. Moreover, of the 20 donors and 16 recipients included for clinical evaluation, only 11 donors and 8 recipients were followed up 1 year post-transplant, which potentially introduced bias into our clinical evaluation, and contributed to a reduction of statistical power to which we could draw our conclusions with. Furthermore, using a sub-cohort for clinical evaluation could potentially introduce selection bias when estimating associations between the choroid and renal function. Therefore, further studies with a larger and more complete dataset would permit a higher confidence regarding the significance of our analyses. However, while the results of the clinical evaluation may suffer from a lack of complete image or clinical data, our primary objective of comparing manual CT against automated CT remain undiluted by these limitations.

Another limitation is that the automated approach does require some, albeit minimal form of manual intervention to select the edge endpoint pixels before tracing each boundary of the choroid in a single B-scan. Another limitation is the possibility of under-approximating the C-S junction for choroids such as the one in figure \ref{fig:major_disc_examples}(b). Finally, while we have provided an analysis on regional quantities of the choroid and their potential links to microvascular injury and renal function, further work would aim to quantify any relationships with the vasculature within the choroid. This could be done for example by computing the choroid vasculature index, which has recently become a popular and more robust way to characterise the choroid for ocular and non-ocular pathology \cite{nakano2020choroid, iovino2020choroidal, betzler2022choroidal, han2022choroidal}. 

\section{Conclusion}
Our automated segmentation methodology can replace human measures for studying the choroid in renal disease, and potentially other clinical conditions. We observed no systematic or proportional differences when comparing manual CT to automated CT measurements. While the automated approach agreed strongly with manual ones, automated CT eliminated angular error and therefore had greater precision when assessing within-patient longitudinal data. The error in manual measurement ultimately led to a lack of strength and significance in their clinical associations. High precision and reproducibility of image measures are critical for longitudinal studies of the choroid where repeated measurements are made using highly sensitive retinal imaging modalities for predicting clinical outcomes.

We observed significant choroidal thickening in recipients 1-year post-transplant and the choroid corresponded to improved eGFR, serum creatinine and urea levels. In donors, we observed choroidal inflation in the first 4 weeks after the unilateral nephrectomy, followed by significant thinning 1-year post-transplant. Initial inflation of the choroid could possibly be related to compensatory renal hypertrophy, while long-term thinning may be linked to long-term impairment of renal function. Future work should aim at a fuller analysis with a larger and more complete dataset in order to increase statistical power.

Our general purpose segmentation methodology is freely available on GitHub \cite{burke2022gpet}, and we are working on producing an automated, end-to-end and open source framework for choroidal image analysis in OCT data. Automated tools to assist clinical prognosis and treatment will strengthen future clinical studies, enabling the delivery of robust, reproducible and responsible research in areas which have previously been at risk of human error.

\section{Acknowledgements}
This work was supported by the Medical Research Council [grant number MR/N013166/1], as part of the Precision Medicine Doctoral Training Programme with the University of Edinburgh. The authors would like to thank all participants in the study as well as all staff at the Edinburgh Imaging Facility who contributed to image acquisition for this study.

\section{Conflicts of Interest}
The authors declare no conflicts of interest.

\bibliographystyle{unsrt}
\bibliography{bibfile.bib}

\onecolumn
 
\setcounter{figure}{0}
\renewcommand{\thefigure}{S\arabic{figure}}
\setcounter{table}{0}
\renewcommand{\thetable}{S\arabic{table}}
\pagebreak

\section*{Supplementary Material}

\begin{figure*}[!b]
    \centering
    \includegraphics[width = 0.8\textwidth]{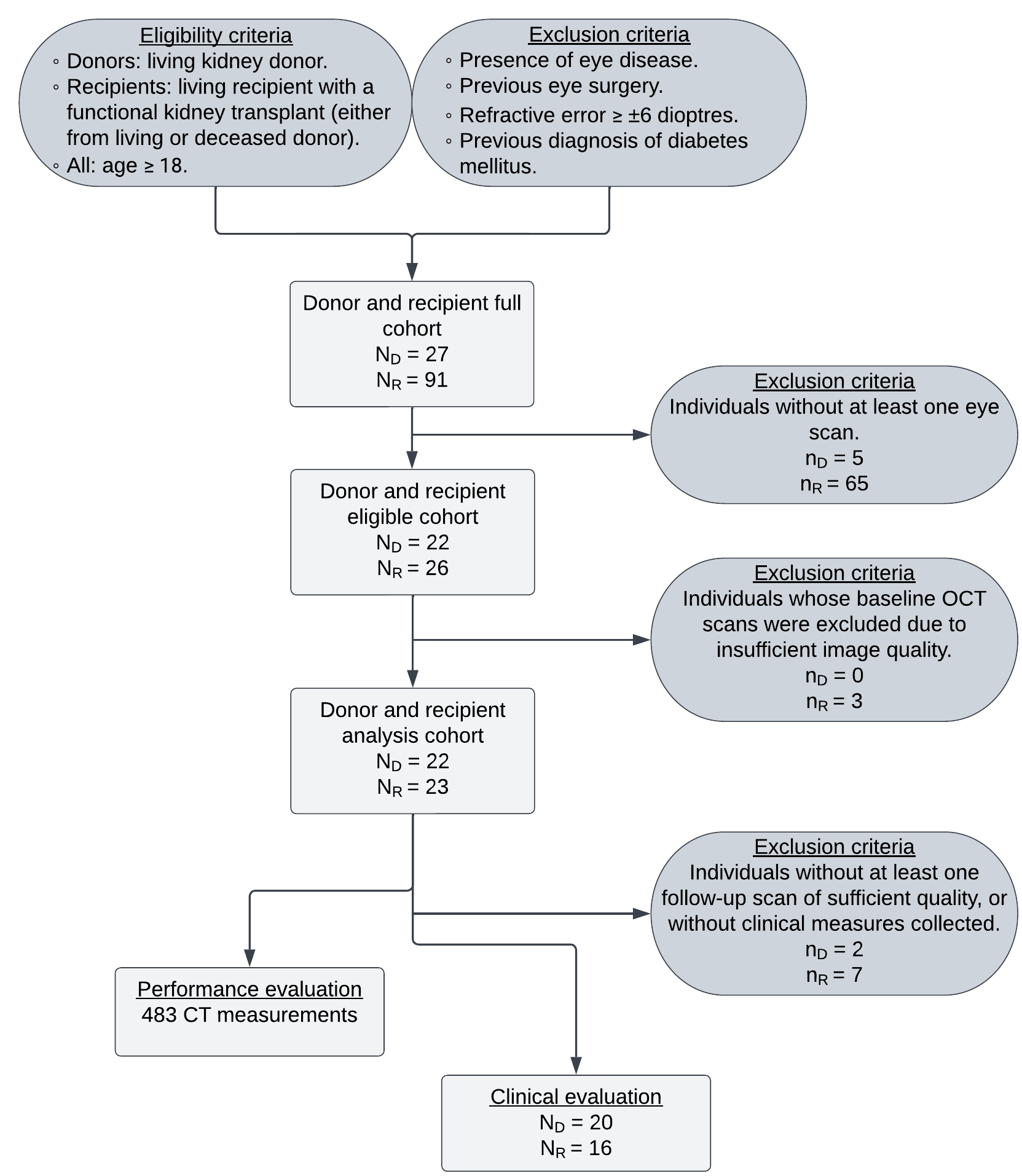}
    \caption{Flow chart visualising donor and recipient cohort specification. $N_D$ and $N_R$ represent number of donors and recipients included at each level. $n_D$ and $n_R$ represent number of donors and recipients excluded at each level, respectively.}
    \label{fig:datasetAB_flow}
\end{figure*}

\begin{figure*}[!tbp]
    \centering
    \includegraphics[width=\textwidth]{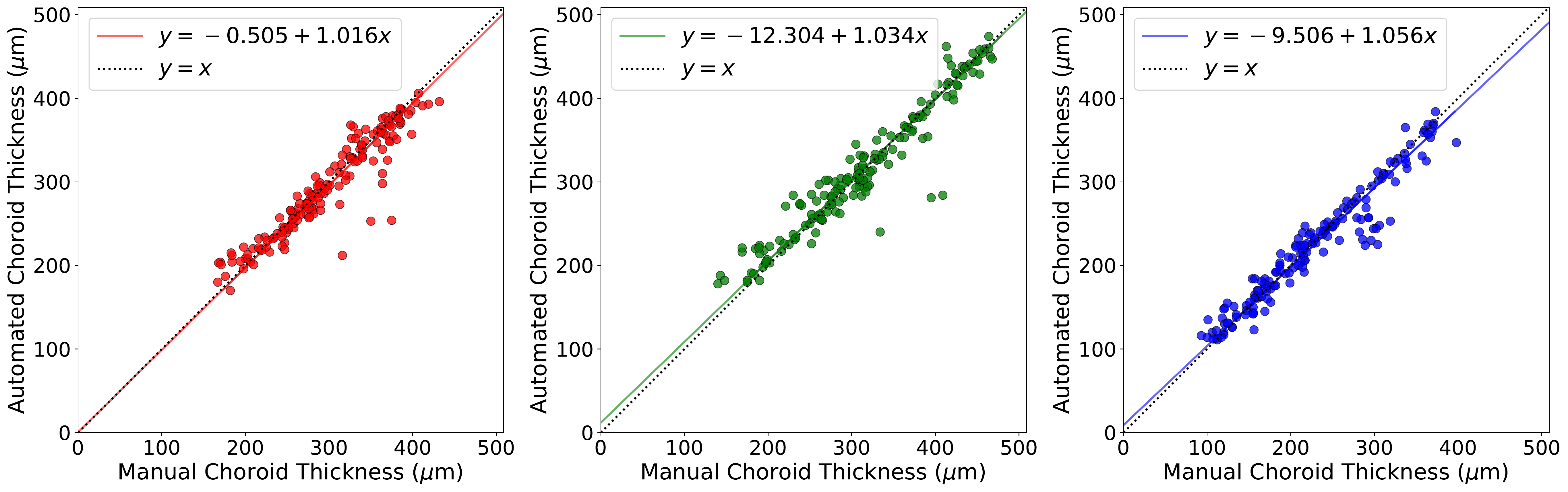}
    \caption{Correlation plot comparing manual and automated choroid thickness measurements, stratified by macular location; temporal (left), subfoveal (middle), nasal (right).}
    \label{fig:corr_plot_loc}
\end{figure*}

\begin{figure*}[!tbp]
    \centering
    \includegraphics[width=0.8\textwidth]{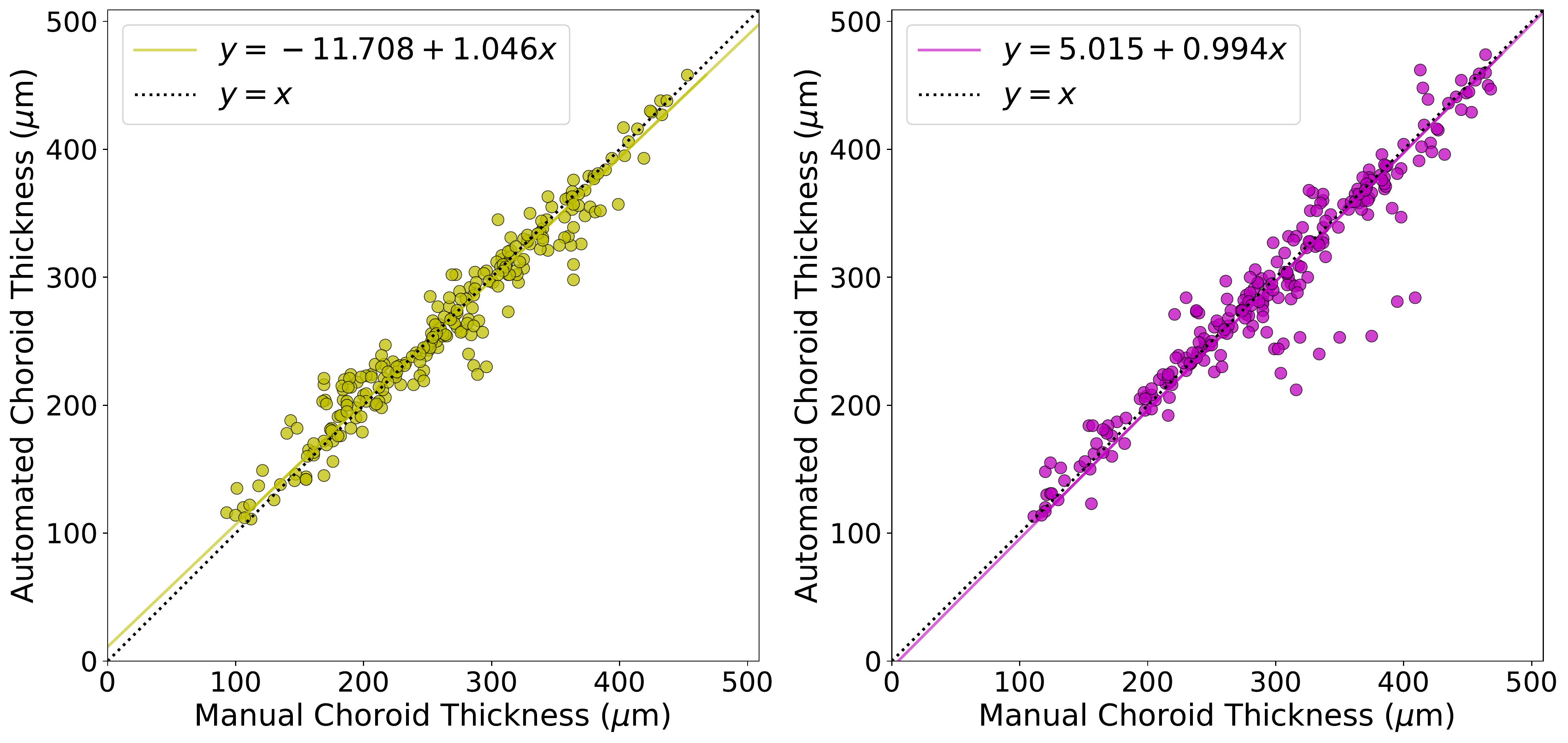}
    \caption{Correlation plot comparing manual and automated choroid thickness measurements, stratified by cohort; donors (left), recipients (right).}
    \label{fig:corr_plot_cohort}
\end{figure*}

\newpage
\begin{table*}[!tbp]
\centering
\begin{tabular}{lllllllllll}
\cmidrule(l){1-4}\cmidrule(l){6-11}
\multicolumn{4}{c}{Pearson, $r_p$} & & \multicolumn{6}{c}{$P$-value, $P$} \\
 \cmidrule(l){1-4}\cmidrule(l){6-11}
 & \multicolumn{3}{c}{Donors} & & \multicolumn{3}{c}{Donors} & \multicolumn{3}{c}{Recipients} \\ \cmidrule(l){2-4}\cmidrule(l){6-8}\cmidrule(l){9-11} 
\multicolumn{1}{l}{Image Measure} & eGFR & Creatinine & Urea & & eGFR & Creatinine & Urea & eGFR & Creatinine & Urea \\
\cmidrule(l){1-4}\cmidrule(l){6-11}
Automated CA & -0.07 & 0.12 & -0.32 & & 0.87 & 0.77 & 0.45 & \textbf{0.04} & \textbf{0.02} & \textbf{0.03} \\
Automated CT & -0.09 & 0.13 & -0.27 & & 0.82 & 0.75 & 0.51 & \textbf{0.01} & \textbf{0.03} & \textbf{0.03} \\
Manual CT & -0.07 & 0.14 & -0.28 & & 0.87 & 0.73 & 0.50 & 0.06 & \textbf{0.03} & \textbf{0.04} \\
\cmidrule(l){1-4}\cmidrule(l){6-11}
\end{tabular}
\caption{(Left) Pearson, $r_p$, correlation analysis between choroidal metrics and eGFR, serum creatinine and urea for the donor cohort. Results presented in this table fail to show evidence of statistical significance. (Right) Corresponding $\textit{P}$-values from Pearson correlation analysis in both cohorts. Bold $\textit{P}$-values represent statistically significant correlations.}
\label{tab:clinic_corr_nonsig}
\end{table*}

\begin{table*}[!t]
\centering
\begin{tabular}{@{}llllllllll@{}}
\toprule
\multicolumn{1}{c}{Response variable $\rightarrow$} &
  \multicolumn{3}{c}{eGFR} &
  \multicolumn{3}{c}{Creatinine} &
  \multicolumn{3}{c}{Urea} \\ \cmidrule(l){1-1}\cmidrule(l){2-4}\cmidrule(l){5-7}\cmidrule(l){8-10} 
\multicolumn{1}{c}{Independent variable $\downarrow$} &
  \multicolumn{1}{c}{$\hat{\beta}$} &
  \multicolumn{1}{c}{95\% CI} &
  \multicolumn{1}{c}{$P$-value} &
  \multicolumn{1}{c}{$\hat{\beta}$} &
  \multicolumn{1}{c}{95\% CI} &
  \multicolumn{1}{c}{$P$-value} &
  \multicolumn{1}{c}{$\hat{\beta}$} &
  \multicolumn{1}{c}{95\% CI} &
  \multicolumn{1}{c}{$P$-value}\\
  \midrule

Intercept & 0.04 & (-0.28, 0.36) & 0.78 & -0.03 & (-0.34, 0.28) & 0.83 & -0.05 & (-0.36, 0.26) & 0.72 \\
Age & -0.34 & (-0.76, 0.08) & 0.10 & 0.07 & (-0.34, 0.47) & 0.73  & 0.08 & (-0.33, 0.48) & 0.69 \\
\textbf{Sex (Male)} & 0.06 & (-0.29, 0.40) & 0.73 & 0.36 & (0.02, 0.69) & \underline{\textbf{0.04}}  & 0.11 & (-0.22, 0.45) & 0.48 \\
Time from Baseline & 3.5e-3 & (-0.24, 0.25) & 0.98 & -0.02 & (-0.25, 0.21) & 0.86  & -0.06 & (0-0.31, 0.18) & 0.60 \\ 
Approx. Refraction & -0.05 & (-0.47, 0.36) & 0.79 & 0.13 & (-0.27, 0.53) & 0.51  & 0.22 & (-0.18, 0.62) & 0.26 \\
Automated CA & 6.7e-4 & (-0.33, 0.33) & 0.99 & -0.11 & (-0.42, 0.21) & 0.49  & -0.07 & (-0.39, 0.24) & 0.64 \\
Patient Std. Dev. & 0.41 & \multicolumn{2}{l}{Cond. $R^2$: 0.27} & 0.45 & \multicolumn{2}{l}{Cond. $R^2$: 0.38} & 0.35 & \multicolumn{2}{l}{Cond. $R^2$: 0.21} \\
\midrule

Intercept & 0.04 & (-0.28, 0.36) & 0.78 & -0.03 & (-0.34, 0.28) & 0.84  & -0.05 & (-0.35, 0.25) & 0.73 \\
Age & -0.34 & (-0.76, 0.36) & 0.10 & 0.06 & (-0.35, 0.46) & 0.78  & 0.07 & (-0.33, 0.47) & 0.72 \\
\textbf{Sex (Male)} & 0.05 & (-0.30, 0.40) & 0.75 & 0.36 & (0.02, 0.70) & \underline{\textbf{0.04}} & 0.12 & (-0.22, 0.45) & 0.47 \\
Time from Baseline & 3.9e-3 & (-0.24, 0.25) & 0.98 & -0.02 & (-0.25, 0.21) & 0.86  & -0.06 & (-0.31, 0.18) & 0.60 \\ 
Approx. Refraction & -0.05 & (-0.47, 0.36) & 0.80 & 0.13 & (-0.27, 0.54) & 0.49  & 0.22 & (-0.17, 0.62) & 0.24 \\
Automated CT & -0.01 & (-0.34, 0.31) & 0.95 & -0.08 & (-0.39, 0.23) & 0.60  & -0.07 & (-0.38, 0.25) & 0.66 \\
Patient Std. Dev. & 0.41 & \multicolumn{2}{l}{Cond. $R^2$: 0.27} & 0.45 & \multicolumn{2}{l}{Cond. $R^2$: 0.37} & 0.34 & \multicolumn{2}{l}{Cond. $R^2$: 0.21} \\ 
\midrule

Intercept & 0.02 & (-0.31, 0.35) & 0.89 & -1.2e-3 & (-0.32, 0.32) & 0.99  & -0.03 & (-0.34, 0.28) & 0.82 \\
Age & 0.33 & (-0.76, 0.09) & 0.12 & 0.04 & (-0.37, 0.44) & 0.84  & 0.06 & (-0.34, 0.46) & 0.75 \\
Sex (Male) & 0.07 & (-0.29, 0.43) & 0.68 & 0.33 & (-0.01, 0.68) & 0.06  & 0.06 & (-0.28, 0.40) & 0.70 \\
Time from Baseline & 6.0e-4 & (-0.24, 0.26) & 0.96 & -0.02 & (-0.25, 0.21) & 0.88  & -0.05 & (-0.30, 0.19) & 0.66 \\ 
Approx. Refraction & -0.09 & (-0.53, 0.36) & 0.68 & 0.19 & (-0.23, 0.61) & 0.34  & 0.28 & (-0.14, 0.69) & 0.17 \\
Manual CT & -5.2e-4 & (-0.34, 0.34) & 0.99 & -0.13 & (-0.45, 0.20) & 0.42  & -0.19 & (-0.51, 0.13) & 0.23 \\
Patient Std. Dev. & 0.43 & \multicolumn{2}{l}{Cond. $R^2$: 0.28} & 0.44 & \multicolumn{2}{l}{Cond. $R^2$: 0.38} & 0.34 & \multicolumn{2}{l}{Cond. $R^2$: 0.23} \\ 
\midrule
\bottomrule

\end{tabular}%
\caption{LME summary tables for predicting eGFR, serum creatinine and serum urea in donors using each choroidal image measurement. No model produced any statistically significant results with the choroid. $\hat{\beta}$ represent standardised model coefficients, with their corresponding 95\% confidence intervals and associated $\textit{P}$-values.}
\label{tab:mixed_lm_donors}
\end{table*}

\end{document}